\documentclass[twocolumn,aps,prl,amsmath,amssymb,color,longbibliography,superscriptaddress]{revtex4-2}
\usepackage{stmaryrd}
\usepackage{bm}
\usepackage{amsmath}
\usepackage{amssymb}
\usepackage{graphicx}
\usepackage{dcolumn}
\usepackage{bm}
\usepackage{tabularx}
\usepackage{diagbox}
\usepackage{adjustbox}
\usepackage{color}
\usepackage{colortbl}

\begin{document}
\begin{titlepage}
\title{Spin-Spiral Enhancement of Ultrafast Light-Polarization-Robust Magnetization}
\author{Yirui Lu}
\affiliation{Beijing Computational Science Research Center, Beijing 100193, China}
\author{Zeyu Jiang}
\email{jiangzy@csrc.ac.cn}
\affiliation{Beijing Computational Science Research Center, Beijing 100193, China}
\author{Bing Huang}
\email{Bing.Huang@csrc.ac.cn}
\affiliation{Beijing Computational Science Research Center, Beijing 100193, China}
\affiliation{Department of Physics, Beijing Normal University, Beijing 100875, China}
\date{\today}

\begin{abstract}
Ultrafast light-driven magnetization, a frontier in quantum magneto-optics, has traditionally relied on circularly polarized lasers to provide external angular momentum. While increasing efforts have aimed to achieve light-polarization-robust (LPR) magnetization that is insensitive to the form of external light excitation, the underlying mechanism remains largely unclear. Here, we establish the symmetry-constrained rule for LPR magnetization in antiferromagnetic systems. Through real-time time-dependent density functional theory calculations, we observe the strong LPR magnetization in spin-spiral magnets and its suppression in collinear antiferromagnets, confirming our theory. Strikingly, laser excitation induces real-space demagnetization, rotation, and oscillation of atomic spins in spin-spiral monolayer NiI$_2$, whereas rotation is largely suppressed in conventional collinear antiferromagnets. Our work reveals a novel microscopic pathway for ultrafast magnetization that is independent of light polarization, paving the way for advanced femtosecond spin control.
\end{abstract}

\maketitle
\draft
\vspace{2mm}
\end{titlepage}

\textcolor{blue}{\textbf{Introduction.}} Ultrafast optics provide a non-thermal means to induce and manipulate magnetization in materials through distinct microscopic mechanisms, enabling angular momentum exchange among charges, spins and phonons on femto- to picosecond timescales~\cite{Siegrist2019,Chen2019SciAdv,Kirilyuk2010,Okyay2020,Bigot2009,Chen2023}. A common strategy employs femtosecond laser pulses to excite an adjacent ferromagnet, generating a spin current that injects angular momentum across the interface into antiferromagnets or nonmagnets~\cite{Geneaux2024,He2023,Chenz2023,Kang2023,Alexey2011}. However, this approach critically depends on precise heterostructure stacking and lattice matching. Likewise, while laser-excited phonons can indirectly mediate magnetization via electronic-structure modification~\cite{Wu2024,Liu2022,YiruiLu2024}, the requirement of selective and coherent phonon-mode excitation on sub-picosecond timescales limits their suitability for ultrafast control.

Alternatively, circularly polarized laser (CPL) pulses can act as an ultrafast, non-contact source of angular momentum, driving net magnetization through the inverse Faraday effect~\cite{Berritta2016,Ziel1965,Kimel2005,Amano2022,Finazzi2013}. As shown in Fig.~1(a), angular momentum of photons is first transferred to electronic orbitals via optical excitation, described by the light-charge coupling term $\hat{\bm{p}}\cdot\bm{A}$. For example, electrons can be excited from the valence $d_{z^2}$ band to the conduction $d_{+1}=d_{xz}+id_{yz}$ band, which carry orbital angular momentum $m_{L} = 0$ and $m_{L} = 1$, respectively. Subsequently, this finite orbital angular momentum is transferred to the spin degree of freedom via the spin-orbit coupling (SOC) Hamiltonian $\bm{L}\cdot\bm{S}$, where electrons relax from the excited $d_{+1}$ state back to the $d_{z^2}$ state, thereby giving rise to a net spin polarization. Previous studies have shown that CPL pulses can reorient spins in ferromagnets, inducing magnetization components transverse to the initial orientation~\cite{Kimel2007,Li2024JPCL}, or induce transient spin polarization in nonmagnetic materials~\cite{Neufeld2023}. However, the induced magnetization is weak and strongly dependent on laser polarization, thereby confining its realization to specific experimental conditions.

\begin{figure}[tbp]
	\includegraphics[width=\columnwidth]{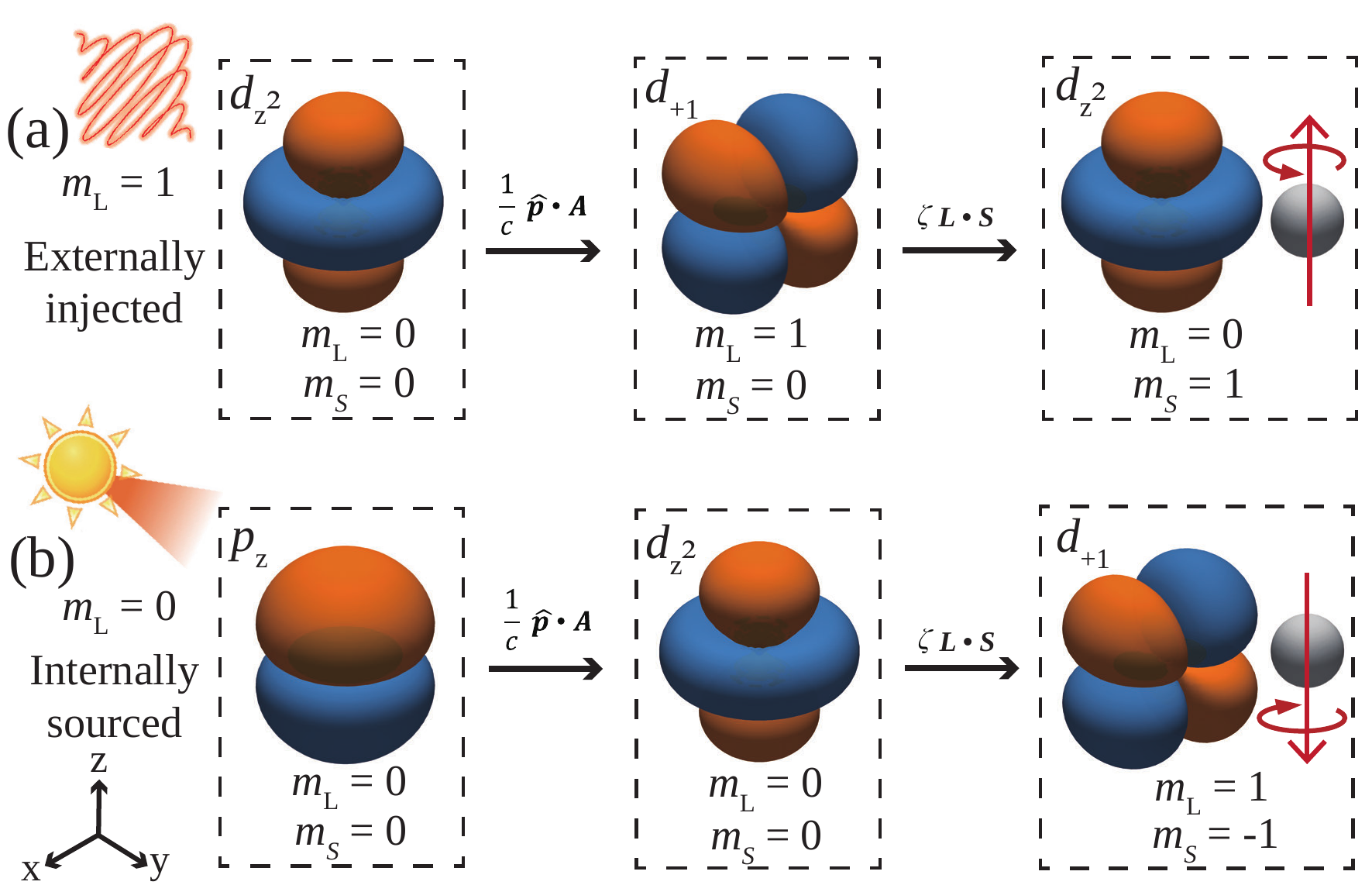}
	\caption{\label{fig:fig1} Schematic of laser-induced magnetization. (a) Magnetization porcess with external photon angular momentum input. (b) Light-polarization-robust magnetization process. In both cases, the incident light first interacts with electronic orbitals through light-charge coupling, with altering (a) and not altering (b) the orbital angular momentum. The orbital angular momentum is then either transferred to (a) or redistributed with (b) the spin degree of freedom through spin-orbit coupling.}
\end{figure}

Notably, all the aforementioned mechanisms involve the injection of external angular momentum, which is first transferred to orbitals or carried by spin currents, and is subsequently converted into spin angular momentum through SOC or exchange interactions. In essence, the angular momentum is exchanged between the external source and the spin system. However, as photospintronic applications increasingly demand broadly tunable and polarization-agnostic approaches for controlling magnetic order~\cite{Wadley2016,Jungwirth2016,Gupta2025,Sui2021}, considerable efforts have been devoted to realizing magnetization in antiferromagnets using linearly polarized light (LPL)~\cite{Zhou2025,Liu2025,Afanasiev2021,Zhao2025,Shen2018,Sato2016,Zhao2025}. To optimize such light-polarization-robust (LPR) magnetization, i.e., those effective for both CPL and LPL, a mechanistic understanding is of both fundamental and technological importance. Here we propose a mechanism that, as illustrated in Fig.~1(b), LPL induces an electronic excitation, for example, from the $p_z$ state to the $d_{z^2}$ state, both of which carry orbital angular momentum $m_{L} = 0$. Then, an intrinsic redistribution of angular momentum occurs without external injection, instead transferring from the orbital to the spin via SOC and ultimately generating net spin polarization with an equal loss of orbital angular momentum. Consequently, optical excitation induces angular momentum exchange between orbital and spin channels via SOC, giving rise to opposite oscillations [Figs.~S1–S2]. Such magnetization originates solely from electronic transitions and remains robust against light polarization. This LPR magnetization process retains the ultrafast timescale and structural simplicity, while removing the need for delicate polarization control.

In this Letter, we show that LPR magnetization in antiferromagnetic systems is governed by the symmetry-constrained rule. Under LPL, most collinear antiferromagnets generate no, or only negligible net magnetization owing to their linear spin degeneracy and sublattice cancellation. By contrast, spin-spiral order intrinsically breaks the symmetry between spin channels, providing an ideal platform for LPR magnetization. We substantiate this by performing real-time time-dependent density functional theory simulations, in which a LPL pulse is applied to collinear antiferromagnets NiPS$_3$ and RuO$_2$, as well as to a cycloidal spin-spiral NiI$_2$ monolayer (see Methods in Supporting Information), the latter exhibiting pronounced angular momentum redistribution between orbital and spin degrees of freedom. We focus on the ultrafast coherent electronic spin dynamics during and immediately following the laser pulse, before significant relaxation and dissipation occur.The emergent magnetization originates from the laser-induced coherence between initially occupied and unoccupied states in the spin-density matrix. During excitation, local moments undergo demagnetization, rotation, and oscillations that collectively yield a macroscopic magnetization.  These results establish a microscopic mechanism for ultrafast LPR magnetization in noncollinear systems and identify spin spirals as a natural setting for femtosecond spin control without external angular momentum input.

\textcolor{blue}{\textbf{Symmetry constraints for LPR magnetization.}}
As the magnetization is insensitive to light polarization, the system symmetry emerges as the most important—if not the only—factor governing LPR behavior. To clarify the conditions enabling LPR magnetization as illustrated in Fig.~1(b), we develop a general understanding on the symmetry constraints in antiferromagnetic systems. The total Hamiltonian under a laser pulse is
\begin{equation}
	\hat{H}(t) = \frac{1}{2} \left[ \hat{\mathbf{p}} + \mathbf{A}_{\mathrm{ext}}(t) \right]^2 + V_{\mathrm{KS}}[\rho(\mathbf{r},t)] + \hat{H}_{\mathrm{SOC}}, \tag{1}
\end{equation}
where a vector potential $\mathbf{A}_{\mathrm{ext}}(t)$ is introduced in the kinetic energy term in the velocity gauge. The second and third terms denote the Kohn–Sham potential and SOC, respectively. By introducing the time-evolution operator $\hat{U}(t)$, the time-dependent density matrix can be expressed as
\begin{equation}
	\hat{\rho}(t) = \hat{U}(t) \hat{\rho}_{0} \hat{U}^\dagger(t), \tag{2}
\end{equation}
with $\hat{U}(t) = \hat{\mathcal{T}} \exp\left(-i \int_0^t \hat{H}(\tau)\, d\tau \right)$, where $\hat{\mathcal{T}}$ is the time-ordering operator and $\hat{\rho}_{0}$ is the ground state density matrix. The time-dependent spin momentum is computed from the trace of the spin-density matrix as
\begin{equation}
	\mathbf{M}(t) = \mu_B \operatorname{Tr}[\hat{\rho}(t) \hat{\boldsymbol{\sigma}}], \tag{3}
\end{equation}
where $\mu_B$ is the Bohr magneton and $\hat{\boldsymbol{\sigma}} = \left(\hat{\sigma}_x, \hat{\sigma}_y, \hat{\sigma}_z\right)$ is the Pauli operator. Therefore, the symmetry of $\hat{\rho}(t)$ becomes the fundamental block governing ultrafast spin dynamics.

Suppose there exists a symmetry operation $\hat{O}$ that commutes with the static Hamiltonian, which includes both spin and atomic degrees of freedom. In principle, a symmetry operation is preserved under laser only if the laser field polarization itself is invariant under that operation. In practice, however, since the laser energy is orders of magnitude smaller than the intrinsic energy scale of the system, symmetry breaking induced by a laser of arbitrary polarization is weak. In addition, the time-dependent Hamiltonian does not respect time-reversal symmetry at any given instant, and the effective time-reversal symmetry recovered through time integration can be broken by a modulated laser field~\cite{Gürtler2004,Friedrich2025,Neufeld2025} Consequently, the only nontrivial symmetry applicable to the spin degrees of freedom is rotation, denoted by $\mathbf{R}$. By inserting the symmetry operator $\hat{O}$ and its inverse into the expression of $\mathbf{M}(t)$, and considering the symmetry properties of the Hamiltonian, we obtain
\begin{equation}
	\begin{aligned}
		\mathbf{M}(t) 
		&= \mu_B \operatorname{Tr}\left[\hat{O} \hat{\rho}(t) \hat{O}^\dagger \cdot \hat{O} \hat{\boldsymbol{\sigma}} \hat{O}^\dagger \right] \\
		&= \mu_B \operatorname{Tr}\left[\hat{\rho}(t) \cdot \hat{O} \hat{\boldsymbol{\sigma}} \hat{O}^\dagger \right] 
		= \mathbf{R}(\varphi) \, \mathbf{M}(t),
	\end{aligned} \tag{4}
\end{equation}
where $\hat{O} \hat{\boldsymbol{\sigma}} \hat{O}^\dagger = \mathbf{R}(\varphi) \hat{\boldsymbol{\sigma}}$ denotes a spin-space rotation acting on the Pauli operators, with rotation angle $\varphi$. We suppose that the rotation axis is along the $z$ axis
\begin{equation} \label{eq:rotation_z}
\begin{bmatrix}
M_x(t) \\
M_y(t) \\
M_z(t)
\end{bmatrix}
=
\begin{bmatrix}
\cos\varphi & -\sin\varphi & 0 \\
\sin\varphi & \cos\varphi  & 0 \\
0          & 0           & 1
\end{bmatrix}
\begin{bmatrix}
M_x(t) \\
M_y(t) \\
M_z(t)
\end{bmatrix}\tag{5}
\end{equation}
Thus, we can obtain
\[
\begin{cases}
M_x(t) = \cos\varphi \, M_x(t) - \sin\varphi \, M_y(t), \\[6pt]
M_y(t) = \sin\varphi \, M_x(t) + \cos\varphi \, M_y(t), \\[6pt]
M_z(t) = M_z(t).
\end{cases}\tag{6}
\]
The general validity of Eq.~(6) under symmetry constraints implies that \textit{magnetization components perpendicular to the rotation axis are forbidden by symmetry, and only the component parallel to the axis can be induced}. Therefore, pronounced LPR magnetization induced by laser is not a general feature of antiferromagnets. Instead, it depends sensitively on the underlying spin textures and symmetries. 

\begin{figure}[tbp]
    \includegraphics[width=\columnwidth]{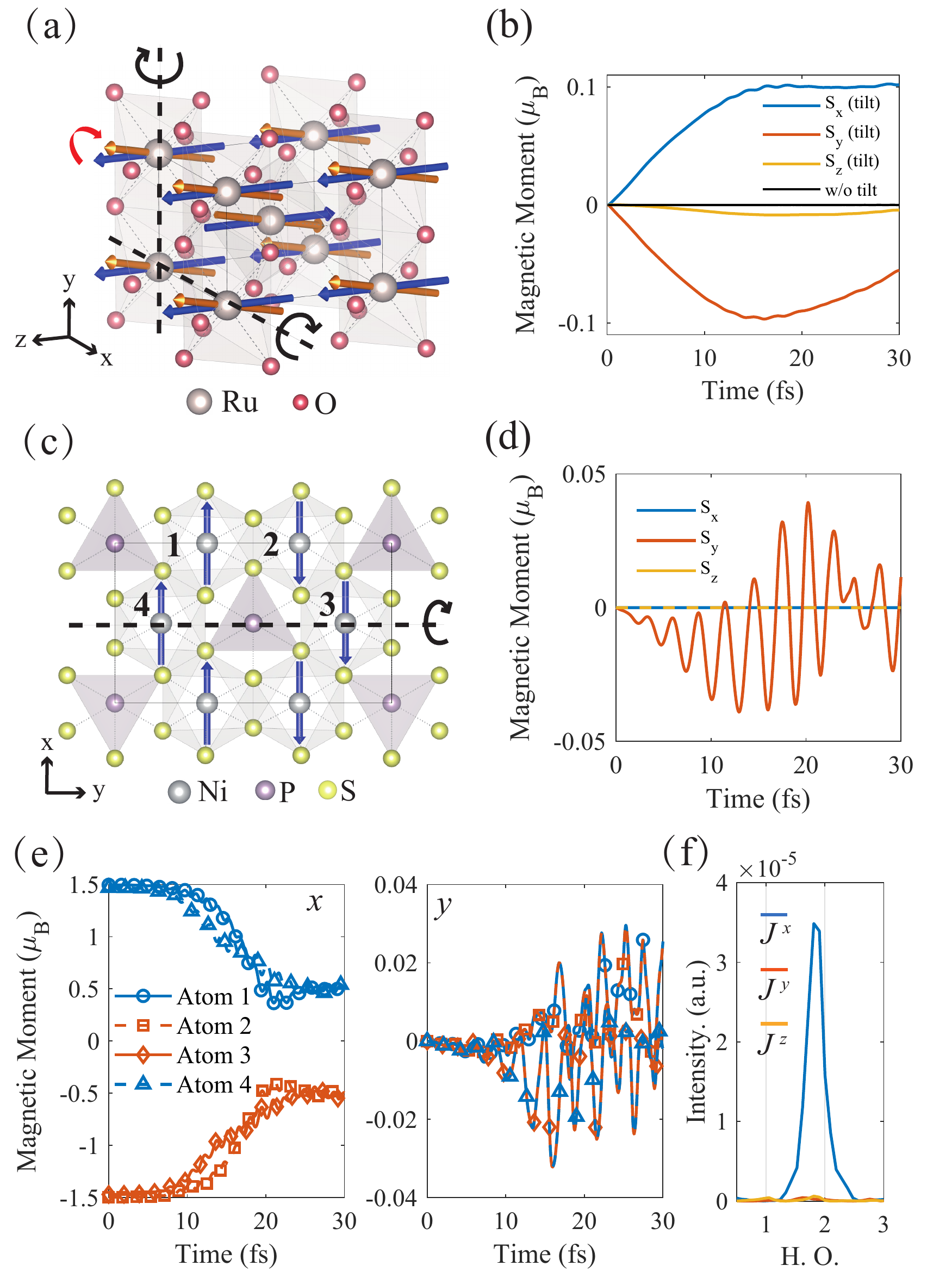}
    \caption{\label{fig:fig2} Laser induced magnetization processes in collinear antiferromagnets.
	(a) Crystal structure of RuO$_2$, with blue and red arrows indicating spins parallel to the $z$ axis and tilted by approximately $20^\circ$, respectively. Two orthogonal rotation axes along $x$ and $y$ directions are indicated by black-dashed lines.
    (b) Time evolution of the total magnetization (per antiferromagnetic unit cell) in different directions under a LPL pulse ($x$-polarized, $\hbar\omega = 1~\text{eV}$) in RuO$_2$ with and without spin tilting. In the static ($t=0$) case, RuO$_2$ has a small ferrimagnetic net moment after tilt, which is set to zero for clarity in the plot.
	(c) Crystal structure of monolayer NiPS$_3$, with spin orientations indicated by blue arrows. 
    (d) Time evolution of the total magnetization under a LPL pulse ($x$-polarized, $\hbar\omega \gtrsim E_{g}$, $E_g$ denotes the band gap) in monolayer NiPS$_3$. 
	(e) Time evolution of the local spin moments of individual atoms along the $x$ and $y$ axes, respectively. The atom indices are indicated in (c). 
	(f) Fourier spectra of spin currents in monolayer NiPS$_3$ polarized along $x$, $y$ and $z$ directions. Harmonic order (H. O.) is defined relative to the input laser frequency.} 
\end{figure}

Multiple symmetry operations generally exist in antiferromagnets. For instance, two rotational operations (possibly followed by a translation) with orthogonal axes strictly forbid any net spin component by symmetry, as the direction parallel to one rotation axis is also perpendicular to the other. As a representative case, we consider the rutile-type tetragonal RuO$_2$ with a collinear antiferromagnetic spin configuration [Fig.~2(a)]. When the spins are aligned along one of the tetragonal axes (e.g., the $z$ axis), other two perpendicular directions serve as rotation axes that mutually constrain magnetization generation, thereby preventing the emergence of net spins in all spatial directions (black line in Fig.~2(b)). Remarkably, tilting the spins away from the $z$ axis breaks these rotational symmetries, allowing a finite net magnetization to appear in three directions [Fig.~2(b)], validating our symmetry-constrained rule. The magnetization magnitude also dependent on the spin tilt angle [Fig.~S3]. Such total symmetry-constrained magnetization are also observed in NiO with specifically designed high-symmetry spin configuration [Fig.~S4]. 

More generally, in collinear antiferromagnets, any nontrivial symmetry operation that exchanges two oppositely oriented spin sublattices is equivalent to an effective $180^\circ$ rotation with an axis perpendicular to the spin direction, thereby confining the symmetry-allowed changes of magnetization solely to the transverse components. In addition, spin variations are primarily driven by laser-induced spin currents~\cite{Krieger2017}, which are typically polarized along the initial spin direction. Thus, transverse spin currents capable of generating symmetry-allowed spin accumulations are very weak, leading to minor magnetization. To demonstrate this, as shown in Fig.~2(c), we take monolayer collinear antiferromagnetic NiPS$_3$, which possesses only one rotational symmetry, as a typical example. Under an $x$-polarized laser, only minor spin fluctuations ($-0.04$ to $0.04~\mu_\mathrm{B}$) appear along $y$ [Fig.~2(d)], as the $C_2$ rotation axis along $y$ forbids magnetization in other directions.

Time evolution of local spins shows that, although demagnetization along $x$ is several orders of magnitude larger than the growth of transverse components along $y$ [Figs.~2(e)], it cancels within each atomic pair (atom 1–2, atom 3–4) based on the symmetry-constrained rule, leaving minor $y$-components originating from weak spin current polarized along $y$ [Fig.~S5]. The second-harmonic peak of the $x$-polarized spin current exceeds the $y$- and $z$-polarized components by over a factor of 50 [Fig.~2(f)], confirming that laser-induced spin current is mainly polarized along initial spin direction. As a result, oppositely oriented local moments are synchronously reduced, yielding no net magnetization, while small residual transverse components are insufficient to produce appreciable LPR magnetization. Interestingly, both pronounced demagnetization and appearance of weak transverse spin components have been observed in various antiferromagnetic and nonmagnetic materials~\cite{Neufeld2023,Elliott2020,Dewhurst2018}. Based on these understanding, to achieve prominent LPR magnetization, two criteria must be satisfied:

\begin{enumerate}
	\item[(i)] \textit{Light-induced magnetization is allowed only along an effective $C_n$ axis, with transverse components forbidden. If two orthogonal $C_n$ axes coexist, it is forbidden in all directions.}
	\item[(ii)] \textit{Temporal growth of net magnetization arises from the accumulation of light-induced spin currents. A large spin current magnitude requires a finite initial spin component along the $C_n$ axis.}
\end{enumerate}

\textcolor{blue}{\textbf{Strong LPR magnetization in NiI$_2$.}}
Helimagnets with noncollinear spin textures exhibit markedly reduced spin-group symmetry, as the spins rotate periodically along the helical axis throughout the lattice, satisfying the criterion (i). In such systems, the initial spins point along multiple directions that are generally incompatible with the symmetry of the underlying atomic lattice, thus allowing substantial spin currents to develop along nearly all spin-polarization directions, satisfying the criterion (ii). Furthermore, the helical magnetic order is often associated with sizeable SOC, which enables efficient angular momentum redistribution between spin and orbital channels and thereby allows substantial spin accumulation even without external angular-momentum injection, i.e., LPR magnetization.

\begin{figure}[tbp]
	\includegraphics[width=\columnwidth]{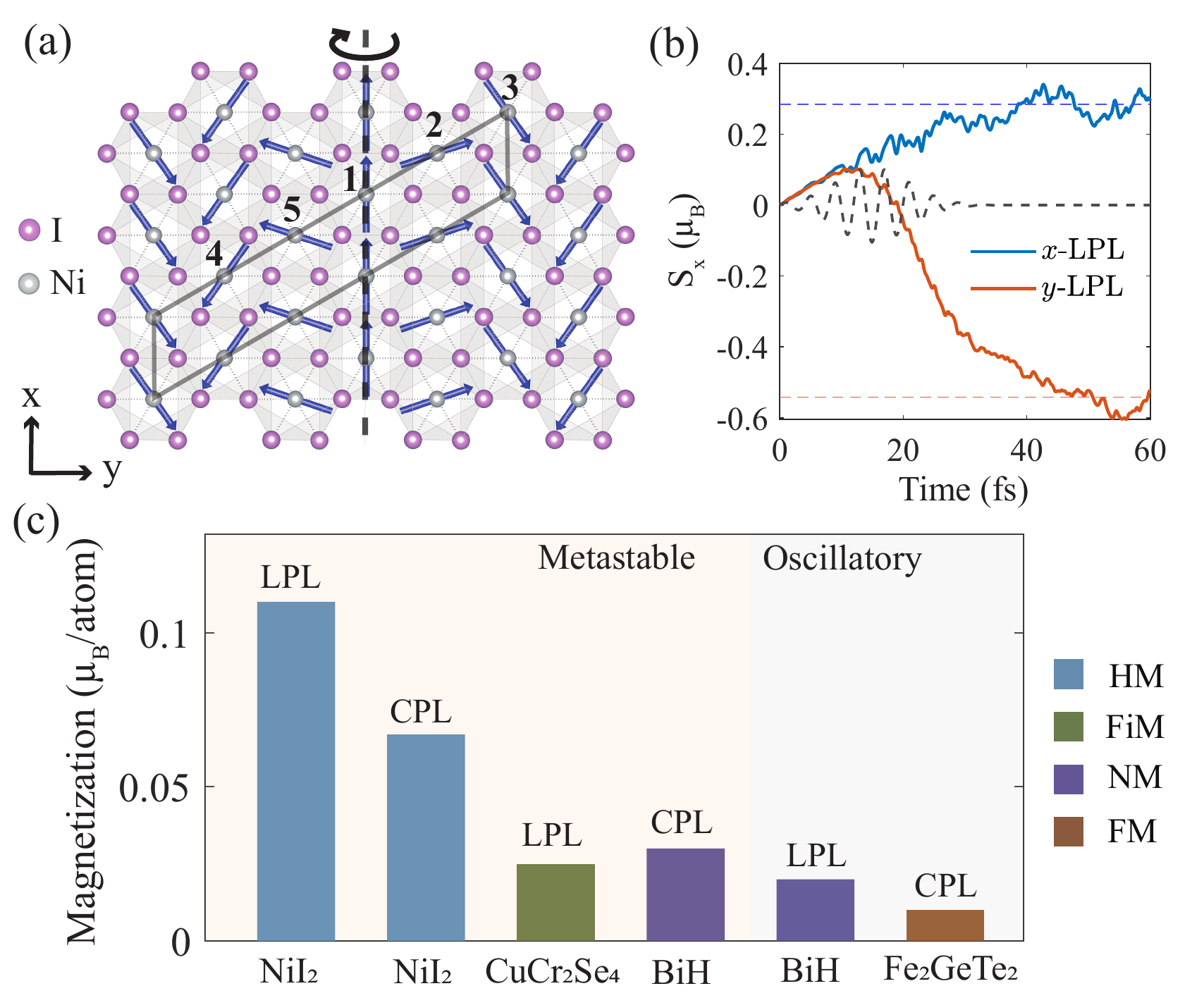}
	\caption{\label{fig:fig3} Laser induced magnetization in NiI$_2$. (a) The spin-spiral magnetic structure, with the $C_2$ axis and atomic order indicated. (b) Time evolution of the total magnetization (per spin-spiral $5\times1$ cell) in monolayer NiI$_2$ under a LPL pulse with frequency $\hbar\omega \gtrsim E{_g}$ ($E_g$ is the band gap, see Fig.~S7) polarized along the $x$- and $y$-directions, with the average value after 40\,fs indicated by a dashed lines. The input laser pulse is represented by black-dashed curve. (c) Comparison of the light-induced magnetization obtained in this work with previous computational results for CuCr$_2$Se$_4$~\cite{Zhao2025}, BiH~\cite{Neufeld2023}, and Fe$_3$GeTe$_2$~\cite{Li2024JPCL}. For inter-material comparison, the total magnetization is averaged over the magnetic atoms. The magnetization is evaluated using a commensurate five-unit-cell spin-spiral structure as a reference. In realistic incommensurate systems, the magnetization is expected to span a tunable range as discussed in the Supplementary Information.} 
\end{figure}
 
NiI$_2$ is a prototypical type-II van der Waals multiferroic that hosts a spin-spiral magnetic ground state~\cite{Song2025}, exemplifying the advantage of spin-spiral order in generating ultrafast LPR magnetization. This work employs a $5 \times 1$ magnetic supercell in subsequent calculations as a commensurate approximation to the incommensurate spin order, consistent with both theoretical results shown in Fig.~S6 and experimental observations~\cite{Miao2025}. The Ni sites are labeled from 1 to 5, capturing a full period of the spin spiral while preserving a twofold rotational symmetry ($C_2$) along the $x$ axis [Fig.~3(a)]. This $C_2$ operation maps sites\,2 and 5 as well as sites\,3 and 4 onto each other, while site~1 lies on the $C_2$ axis and remains invariant. Consequently, NiI$_2$ holds one rotation axis and each atom carries a finite spin component along this axis, satisfying criteria (i)-(ii). 

As illustrated in Fig.~3(b), LPL pulses along the $x$- and $y$-directions induce net magnetization parallel and antiparallel to the ($C_2$-) $x$-axis, respectively. Interestingly, laser induced magnetization oscillation along the $C_2$ axis has also been observed experimentally~\cite{Gao2024}, validating our theory. The magnetization $S_x$ emerges synchronously with the onset of the laser pulse and reaches an oscillatory equilibrium after about 40\,fs, while the $S_y$ and $S_z$ components remain negligible due to symmetry-imposed cancellations [Fig.~S8]. Along a symmetry-allowed direction, the sign of the induced magnetization is not fixed by symmetry alone but is determined by microscopic details of the spin texture and the optical excitation conditions, such as the laser intensity [Fig.~S9]. Further consideration of the influence of atomic motion, out-of-plane spin configurations, and incommensurate effects are discussed in the Supplementary Information [Figs.~S10–S12]. To emphasize the comparatively larger magnitude, we compare the laser-induced magnetization in NiI$_2$ with that reported in other typical magnetic or nonmagnetic materials, as summarized in Fig.~3(c). The detailed discussion can be found in End Matter and Fig. S13.

\begin{figure}[tbp]
	\includegraphics[width=\columnwidth]{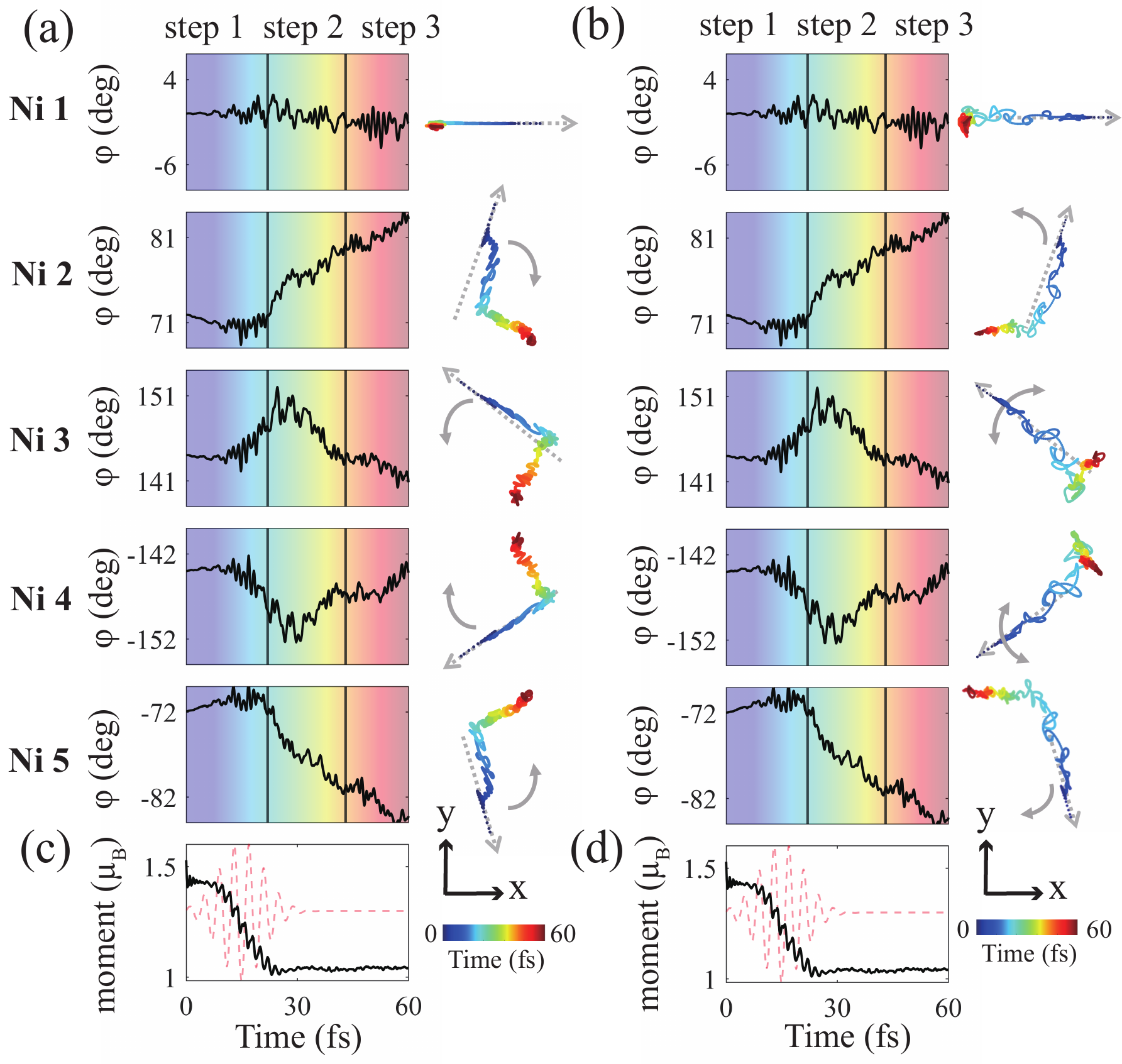}
	\caption{\label{fig:fig4} Local spin dynamics of NiI$_2$ in real space. Time-dependent evolution of local spin moment angles of individual Ni atoms is shown under $x$-polarized (a) and $y$-polarized (b) laser excitation (left panels), together with the corresponding time-resolved trajectories of spin moments (right panels). The gray-dashed arrow indicates the initial spin orientation, and the gray-solid arrow shows the rotation direction, with the $x$-axis defined as $0^\circ$. (c) and (d) show, for $x$- and $y$-polarized laser excitation, respectively, the time evolution of the arithmetic average of the magnitudes of local spin moments over all Ni atoms. The red-dashed curve is the applied laser pulse.}
\end{figure}

It is interesting to explore the dynamics of local spins in a spin-spiral structure, in contrast to the nearly direction-conserving demagnetization observed in collinear antiferromagnets [Fig.~S14]. In Fig.~4, the real-space spin dynamics of individual Ni atoms under $x$- and $y$-polarized laser excitation are examined. The angular evolution reveals three distinct stages: local demagnetization, rotation, and oscillation, as highlighted in the left panels of Figs.~4(a) and 4(b) (steps~1–3). In both cases, the magnetic moment of the Ni\,1 atom, which is aligned parallel to the ($C_2$-) $x$-axis, primarily decreases in magnitude with minimal angular rotation. Meanwhile, the magnetic moments of the other Ni atoms (Ni\,2 to Ni\,5) first decrease in magnitude [Figs.~4(c) and~4(d)], followed by a delayed rotational response lasting approximately 30\,fs, and eventually settle into oscillations around a fixed angle. Although each atom rotates by about $8^{\circ}$, the $C_2$-related pairs (Ni\,2, Ni\,5) and (Ni\,3, Ni\,4) exhibit almose opposite changes, resulting in full cancellation in $y$ and a residual net magnetization along $x$. This is also reflected in the spin trajectories, particularly during the rotation step (highlighted in green and yellow), where opposing rotation directions are marked by gray arrows. Such cancellation arises from the system-wide balance of spin-current-driven local dynamics~\cite{Elliott2020}. The spin trajectories also exhibit distinct behaviors under $x$- and $y$-polarized laser excitation. For instance, Ni\,2 rotates clockwise under $x$-polarization but counterclockwise under $y$-polarization. In addition, Ni\,3 and Ni\,4 rotate toward each other under $x$-polarization, whereas under $y$-polarization they undergo back-and-forth oscillations near their initial orientations. These results highlight that, unlike collinear antiferromagnets where demagnetization occurs almost without transverse spin components, spin rotations are effectively induced in spin-spiral, giving rise to pronounced magnetization.

\textcolor{blue}{\textbf{Conclusion.}}
In summary, we have proposed the mechanism for LPR magnetization, as demonstrated in the prototypical spin-spiral system NiI$_2$. The net magnetic moments can efficiently emerge under excitation from arbitrary light polarization, in contrast to the collinear spin systems. The noncollinear spin arrangement inherently breaks symmetry constraints, enabling laser-induced spin currents along multiple directions. Such a mechanism is absent or weakened in collinear or nonmagnetic systems, thereby limiting their magnetization response. Furthermore, the induced magnetization primarily originates from the off-diagonal spin-density correlation, reflecting inter-state quantum coherence rather than simple occupation changes. These insights identify helimagnets as a promising platform for all-optical spin control, where their unique spin structures enable ultrafast manipulation of magnetic anisotropy in low-dimensional magnetic systems.

\section{Acknowledgments}
We thank the helpful discussions with Drs. Yang Li and Zijing Ding. This work is supported by National Natural Science Foundation of China (NSFC) (Grants Nos. W2511008 and 12088101), Science Challenge Project (Grant No. TZ2025013).

\subsection{End Matter}
\textcolor{blue}{\textbf{Off-diagonal spin-density correlation.}} A microscopic understanding of the light-induced magnetization in NiI$_2$ can be obtained from the occupied time-dependent Bloch states $\lvert \Phi_{nk}(t) \rangle$ as
\begin{equation}
	\mathbf{M}(t) = \mu_B \sum_{n,k} \langle \Phi_{nk}(t) \rvert \hat{\boldsymbol{\sigma}} \lvert \Phi_{nk}(t) \rangle, \tag{7}
\end{equation}
where we can further express $\lvert \Phi_{nk}(t) \rangle$ as a superposition of the eigenstates of the static (unperturbed) Hamiltonian:
\begin{equation}
	\lvert \Phi_{nk}(t) \rangle = \sum_{i} \lvert \psi_{ik} \rangle \langle \psi_{ik} \lvert \Phi_{nk}(t) \rangle = \sum_{i} \alpha_{ink}(t)  \lvert \psi_{ik} \rangle, \tag{8}
\end{equation}
with $\lvert \psi_{ik} \rangle$ denoting the Bloch states prior to the laser pulse. Combining Eqs.~(7) and (8), we have
\begin{equation}
	\begin{aligned}
		\mathbf{M}(t) &= \sum_{i,j} \sum_{k} \left( \sum_{n} \alpha^{*}_{ink}(t) \alpha_{jnk}(t) \right) \langle \psi_{ik} \rvert \hat{\boldsymbol{\sigma}} \lvert \psi_{jk} \rangle \\
		&= \sum_{i,j} \sum_{k} A_{ijk}(t) \langle \psi_{ik} \rvert \hat{\boldsymbol{\sigma}} \lvert \psi_{jk} \rangle,
	\end{aligned} \tag{9}
\end{equation}
where $A_{ijk}(t)$ represents the overlap matrix between the excited states and the unperturbed eigenstates. $A_{ijk}(t)$ is strongly influenced by the external light field, with stronger laser inducing greater overlaps. In contrast, the intrinsic ability of the system to respond to ultrafast laser excitation is primarily encoded in the spin matrix elements $\langle \psi_{ik} \rvert \hat{\boldsymbol{\sigma}} \lvert \psi_{jk} \rangle$.

\begin{figure}[tbp]
	\includegraphics[width=\columnwidth]{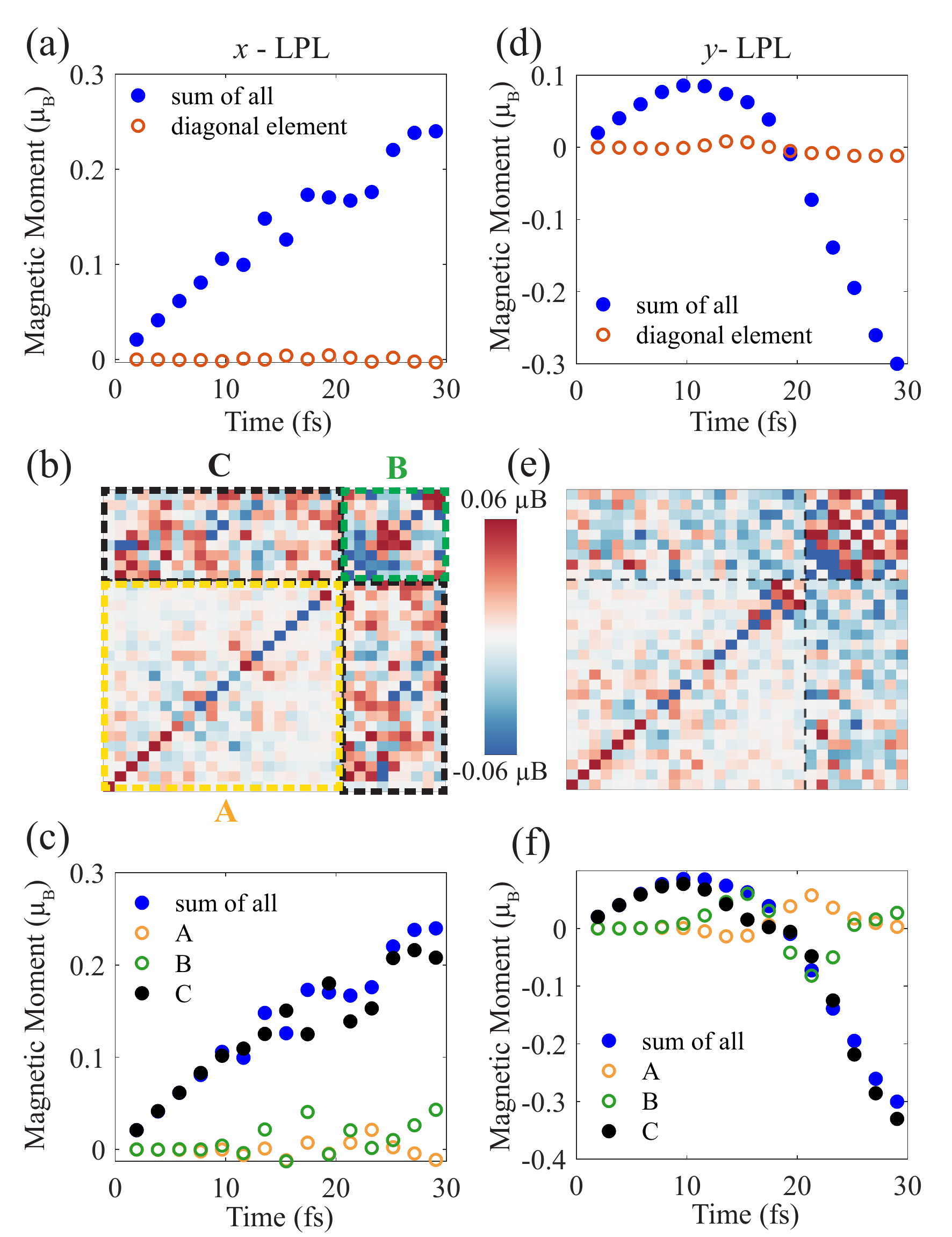}
	\caption{\label{fig:fig5} State-resolved contributions to the net magnetization under  $x$- and $y$-polarized laser excitation in NiI$_2$. (a) Comparison between the diagonal-term contribution and the total magnetization for $x$-polarized light. (b) Matrix representation of $\mathbf{M}_{ij}(t) = \sum_{k} A_{ijk}(t) \langle \psi_{ik} \rvert \hat{\boldsymbol{\sigma}} \lvert \psi_{jk} \rangle$ including both occupied and unoccupied bands at 30\,fs; colored-dashed lines mark the occupation boundary. (c) Partial summation of the block-wise contributions from (b), plotted alongside the total magnetization. (d–f) Corresponding results for $y$-polarized laser excitation. }
\end{figure}

It is natural to decompose $\mathbf{M}(t)$ into diagonal ($i = j$) and off-diagonal ($i \ne j$) contributions, as different $k$-points do not hybridize. The diagonal terms, $A_{iik}(t) = \sum_{n} |\langle \psi_{ik} \lvert \Phi_{nk}(t) \rangle |^2$, represent the occupation change of the state $\lvert \psi_{ik} \rangle$ due to direct transitions, whereas the off-diagonal terms describe the quantum coherence established between different transitions, namely $\lvert nk \rangle \rightarrow \lvert ik \rangle$ and $\lvert nk \rangle \rightarrow \lvert jk \rangle$. However, despite the evident time evolution of occupations [Fig.~S15], the diagonal contributions to the net magnetization remain nearly zero, as shown in Figs.~5(a) and 5(d). 

Given the dominance of off-diagonal contributions, the advantage of spin spirals in generating net magnetization can be understood directly from the spin matrix elements $\langle \psi_{ik} \rvert \hat{\boldsymbol{\sigma}} \lvert \psi_{jk} \rangle$. Although spin is not strictly conserved in the presence of SOC, its energy scale is much smaller than the electronic bandwidth, making spin a useful approximate quantum label. In collinear spin systems, two Bloch states $\lvert \psi_{ik} \rangle$ and $\lvert \psi_{jk} \rangle$ ($i \ne j$) are orthogonal within the same spin channel and spatially separated between opposite spin channels. In both cases, the spin matrix element $\langle \psi_{ik} \rvert \hat{\boldsymbol{\sigma}} \lvert \psi_{jk} \rangle$ largely vanishes, resulting in weak off-diagonal contributions to $\mathbf{M}(t)$. In contrast, in spin-spiral systems, spin is no longer a good quantum number even approximately due to the inherent non-collinearity. As a result, a reasonable balance emerges between the spatial overlap of wavefunctions and their spin non-orthogonality, enabling finite values of $\langle \psi_{ik} \rvert \hat{\boldsymbol{\sigma}} \lvert \psi_{jk} \rangle$ and thus nonvanishing off-diagonal contributions to $\mathbf{M}(t)$. This explains why, as shown in Fig.~3(e), NiI$_2$ exhibit significantly stronger magnetic responses under both linearly and circularly polarized excitation.

To further understand the role of off-diagonal coherence, the band-resolved representation of $\mathbf{M}_{ij}(t) = \sum_{k} A_{ijk}(t) \langle \psi_{ik} \rvert \hat{\boldsymbol{\sigma}} \lvert \psi_{jk} \rangle$ at $t = 30$\,fs is plotted as a two-dimensional matrix for $x$- and $y$-polarized excitations in Figs.~5(b) and 5(e), respectively. Given the boundary between occupied and unoccupied bands, the entire $\mathbf{M}_{ij}(t)$ matrix can be partitioned into three blocks: A, B, and C, corresponding to occupied–occupied, unoccupied–unoccupied, and occupied–unoccupied hybridization, respectively. For each block, its contribution to the net magnetization is summed and then plotted as a function of time in Figs.~5(c) and 5(f). Block~A is nearly diagonal, reflecting the perturbation nature of the light-matter interaction, thus the majority of the charge density---associated with the initially occupied states---remains close to the original diagonal configuration. In contrast, Block~B representing the mixing between unoccupied states is more complex, as these states are all newly excited by the laser pulse. The matrix elements in Block~C are predominantly positive for $x$-polarized light and negative for $y$-polarized light, resulting in magnetic responses of opposite signs. As a result, the magnetization is governed by quantum coherence between occupied and unoccupied bands, i.e., the block C, rather than by occupation changes due to the laser excitation.

\textcolor{blue}{\textbf{Comparison of laser-induced magnetization in different materials.}} As shown in Fig. 3(c), both metastable and oscillatory magnetization responses of different materials are included for comparison. In ferrimagnetic CuCr$_2$Se$_4$~\cite{Zhao2025}, composed of stacked CrSe$_2$ sublayers separated by intercalated Cu atoms [Fig.~S13], the absence of symmetry operations interchanging magnetic sites allows a laser-induced finite magnetization in the excited state, independent of the presence of a nonzero equilibrium magnetization. Despite employing a laser intensity nearly an order of magnitude higher than that used in our study, the resulting spin response remains markedly weak. In nonmagnetic BiH~\cite{Neufeld2023}, spin angular momentum can be introduced either via CPL through the inverse Faraday effect [Fig.~1(a)] or by inducing oscillatory spin-moment dynamics under LPL [Fig.~1(b)]. Even with direct angular momentum injection, however, the resulting magnetization reaches only about one-third of that observed in NiI$_2$. For ferromagnetic Fe$_3$GeTe$_2$~\cite{Li2024JPCL}, CPL induces transverse magnetization that is still considerably small. We attribute the weak magnetization to the lack of initial spin moments along the symmetry-allowed directions, as exemplified by monolayer NiPS$_3$, leading to only minor spin currents. Overall, NiI$_2$ exhibits the strongest spin response under both LPL and CPL, firmly establishing spin-spiral materials as prime candidates for realizing strong LPR magnetization.

\begin{figure}[tbp]
	\includegraphics[width=\columnwidth]{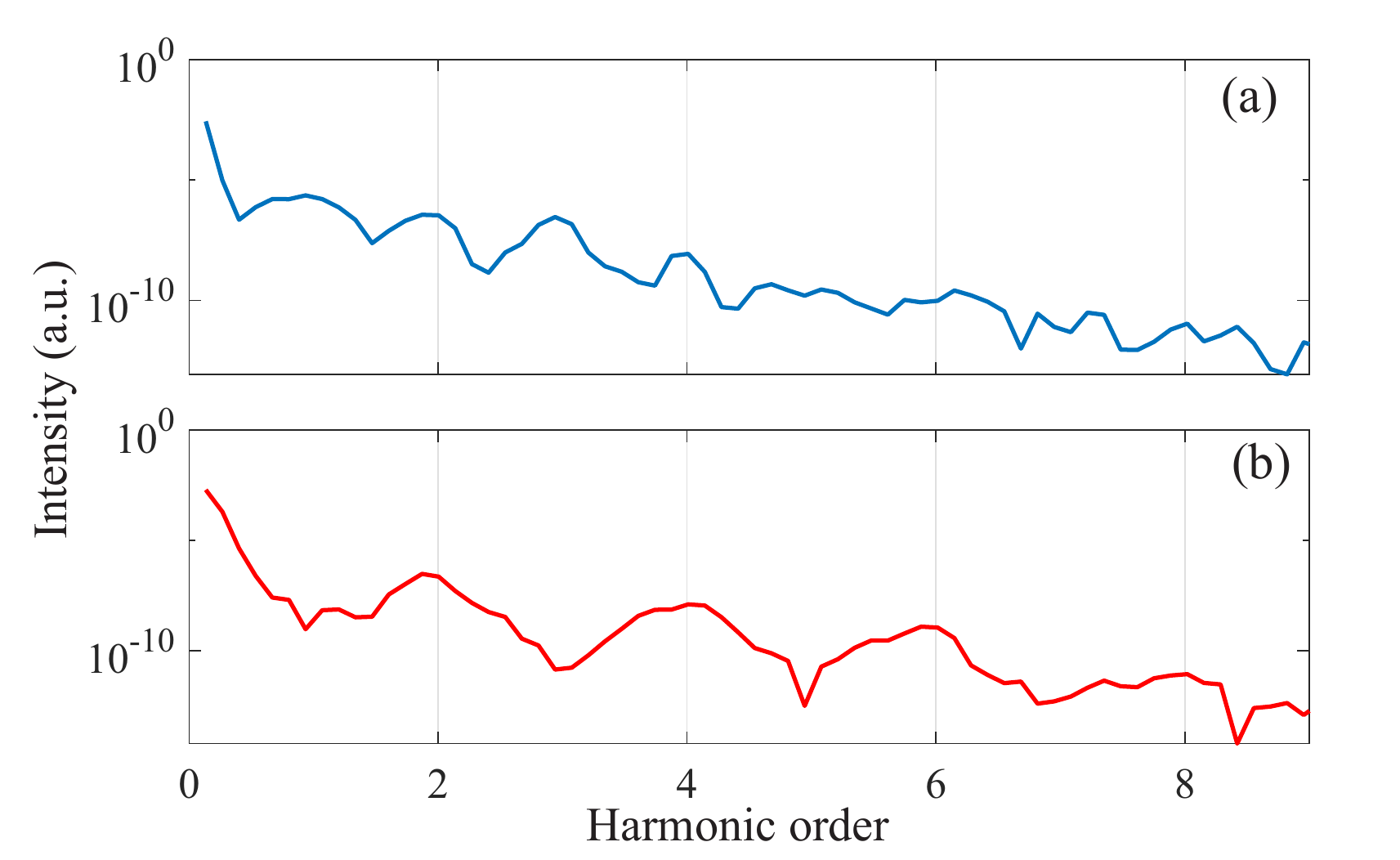}
	\caption{\label{fig:fig6} Fourier spectra of the time-dependent magnetization in the 0–30\,fs interval for laser polarization along the (a) $x$- and (b) $y$-axes, respectively. Harmonic order is defined relative to the driving laser frequency.}
\end{figure}

\textcolor{blue}{\textbf{Harmonic responses of laser-induced magnetization process.}} The Fourier spectra of time evolution of the total magnetization in Fig.~3(b) are presented in Figs.~6(a) and (b), revealing  harmonic responses with progressively reduced intensity at higher orders. For $x$-polarized excitation, both even- and odd-order harmonics appear, whereas only even-order signals are observed for $y$-polarized pulses. This behavior can be understood from the expansion of the spin response:
\begin{equation}
	S^x(t) = S_{0} + \chi_{ix}^{(1)} \boldsymbol{A}(t) + \chi_{iix}^{(2)} \boldsymbol{A}^2(t) + ... (i = x, y), \tag{10}
\end{equation}
where $\boldsymbol{A}$ denotes the vector potential. For $y$-polarized light, the $C_2$ symmetry about the $x$-axis maps $+y$ and $-y$ directions onto each other, suppressing odd-order terms such as $\chi_{yx}^{(1)}$ and $\chi_{yyyx}^{(3)}$, and leaving only even-order contributions. In contrast, $x$-polarized light is not subject to such constraints, allowing all harmonic orders. This symmetry-dependent difference in allowed response channels may also introduce a relative phase, leading to opposite magnetization for $x$- and $y$-polarized pulses.


\begin{thebibliography}{90}%
\makeatletter

\bibitem{Siegrist2019} F. Siegrist, J.A. Gessner, M. Ossiander \textit{et al.},  Light-wave dynamic control of magnetism, Nature \textbf{571}, 240 (2019).

\bibitem{Chen2019SciAdv} Z. Chen and L. W. Wang, Role of initial magnetic disorder: A time-dependent \textit{ab initio} study of ultrafast demagnetization mechanisms, Sci. Adv. \textbf{5}, eaau8000 (2019).

\bibitem{Chen2023} J. Chen, Y. Li, H. Yu \textit{et al.}, Light-induced magnetic phase transition in van der Waals antiferromagnets, Sci. China Phys. Mech. Astron. \textbf{66}, 277511 (2023).

\bibitem{Kirilyuk2010} A. Kirilyuk, A. V. Kimel, and T. Rasing, Ultrafast optical manipulation of magnetic order, Rev. Mod. Phys. \textbf{82}, 2731 (2010).

\bibitem{Okyay2020} M. S. Okyay, A. H. Kulahlioglu, D. Kochan, and N. Park, Resonant amplification of the inverse Faraday effect magnetization dynamics of time reversal symmetric insulators, Phys. Rev. B \textbf{102}, 104304 (2020).

\bibitem{Bigot2009} J. Y. Bigot, M. Vomir, and E. Beaurepaire, Coherent ultrafast magnetism induced by femtosecond laser pulses, Nat. Phys. \textbf{5}, 515 (2009).

\bibitem{Geneaux2024} R. Geneaux, H. Chang, A. Guggenmos, Spin Dynamics across Metallic Layers on the Few-Femtosecond Timescale, Phys. Rev. Lett. \textbf{133}, 106902 (2024).

\bibitem{He2023} J. He, S. Li, T. Frauenheim, and Z. Zhou, Ultrafast Laser Pulse Induced Transient Ferrimagnetic State and Spin Relaxation Dynamics in Two-Dimensional Antiferromagnets, Nano Lett. \textbf{23}, 8348 (2023).

\bibitem{Chenz2023} Z. Chen, J. W. Luo, and L. W. Wang, Light-induced ultrafast spin transport in multilayer metallic films originates from \textit{sp-d} spin exchange coupling, Sci. Adv. \textbf{9}, eadi1618 (2023).

\bibitem{Kang2023} K. Kang, H. Omura, D. Yesudas \textit{et al.}, Spin current driven by ultrafast magnetization of FeRh, Nat. Commun. \textbf{14}, 3619 (2023)

\bibitem{Alexey2011} A. Melnikov, I. Razdolski, O.T. Wehling \textit{et al.}, Ultrafast Transport of Laser-Excited Spin-Polarized Carriers in Au/Fe/MgO(001), Phys. Rev. Lett. \textbf{107}, 076601 (2011)

\bibitem{Wu2024} N. Wu, S. Zhang, D. Chen, Y. Wang, and S. Meng, Three-stage ultrafast demagnetization dynamics in a monolayer ferromagnet, Nat. Commun. \textbf{15}, 2804 (2024).

\bibitem{Liu2022} H. Liu, M. T. Trinh, E. M. Clements \textit{et al.}, Elastically induced magnetization at ultrafast time scales in a chiral helimagnet, Phys. Rev. B \textbf{106}, 035103 (2022).

\bibitem{YiruiLu2024} Y. Lu and Y. Li and B. Huang, Recent progress on photoinduced phase transitions in materials from first-principles calculations, Computational Materials Today \textbf{2-3}, 100012 (2024).

\bibitem{Berritta2016} M. Berritta, R. Mondal, K. Carva, and P. M. Oppeneer, \textit{Ab Initio} Theory of Coherent Laser-Induced Magnetization in Metals, Phys. Rev. Lett. \textbf{117}, 137203 (2016).

\bibitem{Ziel1965} J. P. van der Ziel, P. S. Pershan, L. D. Malmstrom, Optically-Induced Magnetization Resulting from the Inverse Faraday Effect, Phys. Rev. Lett. \textbf{15}, 190 (1965).

\bibitem{Kimel2005} A. V. Kimel, A. Kirilyuk, P. Usachev \textit{et al.}, Ultrafast non-thermal control of magnetization by instantaneous photomagnetic pulses, Nature \textbf{435}, 655 (2005).

\bibitem{Amano2022} T. Amano, Y. Kawakami, H. Itoh \textit{et al.}, Light-induced magnetization driven by interorbital charge motion in the spin-orbit assisted Mott insulator {$\alpha$-RuCl$_3$}, Phys. Rev. Lett. \textbf{14}, L032032 (2022).

\bibitem{Finazzi2013} M. Finazzi, M. Savoini, A. R. Khorsand \textit{et al.}, Laser-Induced Magnetic Nanostructures with Tunable Topological Properties, Phys. Rev. Lett. \textbf{110}, 177205 (2013).

\bibitem{Kimel2007} A. V. Kimel, A. Kirilyuk, F. Hansteen, R. V. Pisarev, and T. Rasing, Nonthermal optical control of magnetism and ultrafast laser-induced spin dynamics in solids, J. Phys.: Condens. Matter \textbf{19}, 043201 (2007).

\bibitem{Li2024JPCL} S. Li, R. Wang, T. Frauenheim, and J. He, Optical-Helicity-Dependent Orbital and Spin Dynamics in Two-Dimensional Ferromagnets, J. Phys. Chem. Lett. \textbf{15}, 5939 (2024).

\bibitem{Neufeld2023} O. Neufeld, N. Tancogne-Dejean, U. De Giovannini, H. Hübener, and A. Rubio, Attosecond magnetization dynamics in non-magnetic materials driven by intense femtosecond lasers, npj Comput. Mater. \textbf{9}, 39 (2023).

\bibitem{Wadley2016} P. Wadley \textit{et al.}, Electrical switching of an antiferromagnet, Science \textbf{351}, 587 (2016).

\bibitem{Jungwirth2016} T. Jungwirth, X. Marti, P. Wadley, and J. Wunderlich, Antiferromagnetic spintronics, Nat. Nanotechnol. \textbf{11}, 231 (2016).

\bibitem{Gupta2025} D. Gupta, M. Pankratova, M.Riepp \textit{et al.}, Tuning ultrafast demagnetization with ultrashort spin polarized currents in multi-sublattice ferrimagnets, Nat. Commun. \textbf{16}, 3097 (2025).

\bibitem{Sui2021} X. Sui, J. Wang, C. Yam, and B. Huang, Two-Dimensional Magnetic Anionic Electrons in Electrides: Generation and Manipulation, Nano Lett. \textbf{21}, 3813 (2021).

\bibitem{Zhou2025} J. Zhou, C. Zhang, Contrasting Light-Induced Spin Torque in Antiferromagnetic and Altermagnetic Systems, Phys. Rev. Lett. \textbf{134}, 176902 (2025)

\bibitem{Liu2025} Y. Liu, S. Guo, Y. Li, C. Liu, Two-Dimensional Fully Compensated Ferrimagnetism,  Phys. Rev. Lett. \textbf{134}, 116703 (2025)

\bibitem{Afanasiev2021} D. Afanasiev \textit{et al.}, Controlling the anisotropy of a van der Waals antiferromagnet with light, Sci. Adv. \textbf{7}, eabf3096 (2021).

\bibitem{Shen2018} L. Q. Shen, L. F. Zhou, J. Y. Shi, M. Tang, Z. Zheng, D. Wu, S. M. Zhou, L. Y. Chen, and H. B. Zhao, Dominant role of inverse Cotton-Mouton effect in ultrafast stimulation of magnetization precession in undoped yttrium iron garnet films by 400-nm laser pulses, Phys. Rev. B \textbf{97}, 224430 (2018).

\bibitem{Sato2016} M. Sato, S. Takayoshi, T. Oka, Laser-Driven Multiferroics and Ultrafast Spin Current Generation, Phys. Rev. Lett. \textbf{117}, 147202 (2016).

\bibitem{Zhao2025} J. Zhao, Y. Feng, K. Dou, X. Li, Y. Dai, B. Huang, and Y. Ma, Light-Induced Spin Slanting in 2D Multiferroic Magnet, ACS Nano \textbf{19}, 24005 (2025).

\bibitem{Gürtler2004} A. Gürtler, F. Robicheaux, W. J. van der Zande and L. D. Noordam, Asymmetry in the Strong-Field Ionization of Rydberg Atoms by Few-Cycle Pulses, Phys. Rev. Lett. \textbf{92}, 033002 (2004).

\bibitem{Friedrich2025} F. Friedrich, P. Herrmann, S.S. Shanbhag, et al. Measurement of optically induced broken time-reversal symmetry in atomically thin crystals. Nat. Photon. (2025).

\bibitem{Neufeld2025} Ofer Neufeld, Linearly Polarized Few-Cycle Pulses Drive Carrier-Envelope Phase-Sensitive Coherent Magnetization Injection, Phys. Rev. Lett. \textbf{135}, 206902 (2025).

\bibitem{Krieger2017} K. Krieger, P. Elliott, T. Müller \textit{et al.}, Ultrafast demagnetization in bulk versus thin films: an \textit{ab initio} study, Journal of Physics: Condensed Matter, \textbf{29}, 224001 (2017)

\bibitem{Elliott2020} P. Elliott, N. Singh, K. Krieger, E.K.U. Gross, S. Sharma, and J.K. Dewhurst, The microscopic origin of spin-orbit mediated spin-flips, Journal of Magnetism and Magnetic Materials, \textbf{502}, 166473 (2020)

\bibitem{Dewhurst2018} J. K. Dewhurst, P. Elliott, S. Shallcross, E. K. U. Gross, and S. Sharma, Laser-induced intersite spin transfer, Nano Lett. \textbf{18}, 1842 (2018).

\bibitem{Song2025} Song, Q., Stavrić, S., Barone, P. \textit{et al.}. Electrical switching of a \textit{p}-wave magnet, Nature \textbf{642}, 64–70 (2025).

\bibitem{Miao2025} M. Miao, N. Liu, W. Zhang, J. Zhou, D. Wang, C. Wang, W. Ji, and Y. Fu, Spin-resolved imaging of atomic-scale helimagnetism in mono- and bilayer $NiI_2$, Proc. Natl. Acad. Sci. U.S.A. \textbf{122}, e2422868122 (2025).

\bibitem{Gao2024} F. Y. Gao \textit{et al.}, Giant chiral magnetoelectric oscillations in a van der Waals multiferroic, Nature \textbf{632}, 273 (2024).

\end{thebibliography}
\end{document}